\documentclass[
]{jpsj2}

\title{Investigation of transition frequencies of two acoustically coupled 
bubbles using a direct numerical simulation technique}

\author{Masato \textsc{Ida}
\thanks{Present address: Center for Promotion of Computational Science and 
Engineering, Japan Atomic Energy Research Institute, 6-9-3 Higashi-Ueno, 
Taito-ku, Tokyo 110-0015; E-Mail: ida@koma.jaeri.go.jp.}}

\inst{Collaborative Research Center of Frontier Simulation Software for Industrial Science, Institute of Industrial Science, University of Tokyo, 
4-6-1 Komaba, Meguro-Ku, Tokyo 153--8505}

\abst{
The theoretical results regarding the ``transition frequencies'' of two 
acoustically interacting bubbles have been verified numerically. The theory 
provided by Ida [Phys.~Lett.~A \textbf{297} (2002) 210] predicted the 
existence of three transition frequencies per bubble, each of which has the 
phase difference of $\pi /2$ between a bubble's pulsation and the external 
sound field, while previous theories predicted only two natural frequencies 
which cause such phase shifts. Namely, two of the three transition 
frequencies correspond to the natural frequencies, while the remaining does 
not. In a subsequent paper [M.~Ida, Phys.~Rev.~E \textbf{67} (2003) 056617], 
it was shown theoretically that transition frequencies other than the 
natural frequencies may cause the sign reversal of the secondary Bjerknes 
force acting between pulsating bubbles. In the present study, we employ a 
direct numerical simulation technique that uses the compressible 
Navier--Stokes equations with a surface-tension term as the governing equations 
to investigate the transition frequencies of two coupled bubbles by 
observing their pulsation amplitudes and directions of translational motion, 
both of which change as the driving frequency changes. The numerical results 
reproduce the recent theoretical predictions, validating the existence of 
the transition frequencies not corresponding to the natural frequency.
}

\kword{bubble dynamics, secondary Bjerknes force, direct numerical 
simulation, natural frequency, transition frequency}

\begin{document}

\maketitle

\section{Introduction}
The secondary Bjerknes force is an interaction force acting between 
pulsating gas bubbles in an acoustic field.\cite{ref1,ref2,ref3} The classical theory 
originated by Bjerknes predicts either attraction only or repulsion only, 
depending on whether the driving frequency stays outside or inside, 
respectively, the frequency region between the partial (or monopole) natural 
frequencies of two bubbles. However, recent studies show that the force 
sometimes reverses its own direction as the distance between the bubbles 
changes.\cite{ref4,ref5,ref6} The first theoretical study on this subject was 
performed by Zabolotskaya.\cite{ref4} Employing a linear coupled oscillator 
model, she showed 
that the radiative interaction between the bubbles, which results in the 
change in the natural frequencies of the bubbles, could cause this reversal. 
In the mid-1990s, Doinikov and Zavtrak arrived at the same conclusion by 
employing a linear theoretical model in which the multiple scattering 
between the bubbles is more rigorously taken into account.\cite{ref5} These 
theoretical results are considered to explain the stable structure formation 
of bubbles in a sound field, called ``bubble cluster'' or ``bubble grape,'' 
which has been observed experimentally by several researchers in different 
fields.\cite{ref7,ref8,ref9,ref10,ref11,ref3}

In both of the theoretical studies mentioned above, it was assumed that the 
reversal is due to the change in the natural (or the resonance) frequencies 
of bubbles, caused by the radiative interaction between bubbles. However, 
those authors had differing interpretations of how the natural frequencies 
change. The theoretical formula for the natural frequencies, used by 
Zabolotskaya\cite{ref4} and given previously by Shima,\cite{ref12} shows that 
the higher and the lower natural frequencies (converging to the partial natural 
frequencies of a smaller and a larger bubble, respectively, when the distance 
between the bubbles is infinite) reveal an upward and a downward shift, 
respectively, as the bubbles come closer to one another.\cite{ref12,ref13} In 
contrast, Doinikov and Zavtrak assumed intuitively that both the natural 
frequencies rise.\cite{ref5,ref6} This assumption seems to explain well the sign 
reversal occurring not only when both bubbles are larger than the resonance 
size but also when one bubble is larger and the other is smaller than the 
resonance size. The sign reversal in the latter case, for instance, is 
thought to occur when the resonance frequency of a larger bubble, increasing 
as the bubbles come closer to one another, surpasses the driving frequency, 
resulting in the change in the pulsation phase of the larger bubble, and 
leading to the change in the phase difference between the bubbles. However, 
this assumption is obviously inconsistent with the previous theoretical 
result for the natural frequencies.\cite{ref12,ref4}

Recently, Ida\cite{ref14} proposed an alternative theoretical explanation for 
this phenomenon, also using the linear model Zabolotskaya used. He claimed that 
this phenomenon cannot be interpreted by only observing the natural 
frequencies, and that it is useful to define the \textit{transition frequencies} that make the phase 
difference between a bubble's pulsation and an external sound $\pi /2$ (or 
$3\pi /2)$. It has been pointed out theoretically that the maximum number of 
natural frequencies and that of transition frequencies are, in general, not 
in agreement in multibubble cases,\cite{ref13,ref15} while they are, as is well 
known, consistent in single-bubble cases where the phase difference between 
a bubble's pulsation and an external sound becomes $\pi /2$ only at its 
natural frequency. (This is not true in strongly nonlinear cases, in which 
the phase reversal can take place even in frequency regions far from 
the bubble's natural frequency; see, \textit{e.g}., refs.~\citen{ref16,ref17}.) 
In a double-bubble case, for instance, that theory predicts 
three transition frequencies per bubble, two of which correspond to the 
natural frequencies.\cite{ref13} A preliminary discussion on a $N$-bubble system\cite{ref15} showed that a bubble in this system has up to $2N-1$ 
transition frequencies, only $N$ ones of which correspond to the natural 
frequencies. More specifically, the number of transition frequencies is in 
general larger than that of natural frequencies. The transition frequencies 
not corresponding to the natural frequency have differing physical meanings 
from those of the natural frequencies; they do not cause the resonance response 
of the bubbles.\cite{ref13,ref15} The theory for the sign reversal of the 
force, constructed based on the transition frequencies, predicts that the sign 
reversal takes place around those frequencies, not around the natural 
frequencies, and can explain the sign reversal in both cases mentioned 
above.\cite{ref14} Moreover, the theory does not contradict the theory for the 
natural frequencies described previously, because all the natural frequencies 
are included in the transition frequencies.

The aim of this paper is to verify the theoretical prediction of the 
transition frequencies by direct numerical simulation (DNS). In a recent 
paper,\cite{ref18} Ida proposed a DNS technique, based on a hybrid advection 
scheme,\cite{ref19} a multi-time-step integration technique,\cite{ref18} and 
the Cubic-Interpolated Propagation/Combined, Unified Procedure (CIP--CUP) 
algorithm,\cite{ref20,ref21} which technique allows us to compute 
the dynamics (pulsation and translational motion) of deformable bubbles in a 
viscous liquid even when the separation distance between the bubbles is 
small.\cite{ref18,ref22} In that DNS technique, the compressible Navier--Stokes 
equations with a surface-tension term are selected as the governing equations, 
and the 
convection, the acoustic, and the surface-tension terms, respectively, in 
these equations are solved by an explicit advection scheme employing both 
interpolation and extrapolation functions which realizes a discontinuous 
description of interfaces between different materials,\cite{ref19} by the 
Combined, Unified Procedure (CUP) being an implicit finite difference method 
for all-Mach-number flows,\cite{ref20,ref21} and by the Continuum Surface 
Force (CSF) model being a finite difference solver for the surface-tension term 
as a volume force.\cite{ref23} Efficient and accurate time integration of the 
compressible Navier--Stokes equations under a low-Mach-number condition is 
achieved by the multi-time-step integration technique,\cite{ref18} 
which solves the different-nature terms in these equations with different 
time steps. Further details of this DNS technique can be found in 
refs.~\citen{ref18,ref19}. Employing this DNS technique, in the present study 
we perform numerical 
experiments involving two acoustically coupled bubbles in order to 
investigate the recent theories by observing the bubbles' pulsation 
amplitudes and the directions of their translational motion. In particular, we 
focus our attention on the existence of the transition frequencies that do not 
cause the resonance response.

The following sections are organized as follows. In \S \ref{sec2}, the 
previously expounded theories are reviewed and reexamined, including those for 
the transition 
frequencies and the sign reversal, and in \S \ref{sec3}, the numerical results 
and a discussion are provided. Section \ref{sec4} presents concluding remarks.

\section{Theories}
\label{sec2}
\subsection{A single-bubble problem}
It is well known that, when the wavelength of an external sound is 
sufficiently large compared to the radius of a bubble (named ``bubble 1,'' 
immersed in a liquid) and the sphericity of the bubble is maintained, the 
following second-order differential equation\cite{ref3,ref24} describes the linear 
pulsation of the bubble:
\begin{equation}
\label{eq1}
\ddot {e}_1 +\omega _{10}^2 e_1 +\delta _1 \dot {e}_1 =-\frac{p_{\rm{ex}} 
}{\rho _0 R_{10} },
\end{equation}
\[
\omega _{10} =\sqrt {\frac{1}{\rho _0 R_{10}^2 }\left[ {3\kappa P_0 
+(3\kappa -1)\frac{2\sigma }{R_{10} }} \right]} ,
\]
where it is assumed that the bubble's time-dependent radius can be 
represented by $R_1 =R_{10} +e_1 $ ($\left| {e_1 } \right|\ll R_{10} )$ with 
$R_{10} $ being its equilibrium radius and $e_1 $ being the deviation in the 
radius, and $\omega _{10} $ is the bubble's natural frequency, $\delta _1 $ 
is the damping factor,\cite{ref25,ref26} $p_{\rm{ex}} $ is the sound pressure at 
the bubble position, $\rho _0 $ is the equilibrium density of the liquid, 
$P_0 $ is the static pressure, $\kappa $ is the polytropic exponent of the 
gas inside the bubble, $\sigma $ is the surface tension, and the over dots 
denote the derivation with respect to time. Assuming $p_{\rm{ex}} =-P_a 
\sin \omega t$ (where $P_a $ is a positive constant), the harmonic 
steady-state solution of eq.~(\ref{eq1}) is determined as
\[
e_1 =K_{S1} \sin (\omega t-\phi _{S1} ),
\]
where
\[
K_{S1} =\frac{P_a }{\rho _0 R_{10} }\sqrt {\frac{1}{(\omega _{10}^2 -\omega 
^2)^2+\delta _1^2 \omega ^2}} ,
\]
\[
\phi _{S1} =\tan ^{-1}\left( {\frac{\delta _1 \omega }{\omega _{10}^2 
-\omega ^2}} \right).
\]
This result reveals that the phase reversal (or the phase difference of 
$\phi _{S1} =\pi /2)$ appears at only $\omega =\omega _{10} $. Moreover, if 
$\delta _1 \ll \omega _{10} $, the bubble's resonance response can take 
place at almost the same frequency. Though the resonance frequency shifts 
away from $\omega _{10} $ as $\delta _1 $ increases, it can in many cases be 
assumed to be almost the same as $\omega _{10} $. (Also, the nonlinearity in 
bubble pulsation is known to alter a bubble's resonance 
frequency.\cite{ref2,ref3,ref27})

\subsection{A double-bubble problem}
When one other bubble (bubble 2) exists, the pulsation of the previous 
bubble is driven by not only the external sound but also the sound wave that 
bubble 2 radiates. Assuming that the surrounding liquid is incompressible 
and the time-dependent radius of bubble 2 can be represented by $R_2 =R_{20} 
+e_2 $ ($\left| {e_2 } \right|\ll R_{20} )$, the radiated pressure field 
is found to be,\cite{ref15}
\[
p(r,t)\approx \frac{\rho _0 R_{20}^2 }{r}\ddot {e}_2 ,
\]
where $r$ is the distance measured from the center of bubble 2. The total 
sound pressure at the position of bubble 1 ($p_{\rm{d1}} )$ is, thus,
\begin{equation}
\label{eq2}
p_{\rm{d1}} \approx p_{\rm{ex}} +\frac{\rho _0 R_{20}^2 }{D}\ddot {e}_2 
,
\end{equation}
where $D$ is the distance between the centers of the bubbles. This total 
pressure drives the pulsation of bubble 1.

Replacing $p_{\rm{ex}} $ in eq.~(\ref{eq1}) with $p_{\rm{d1}} $ yields the 
modified equation for bubble 1,
\begin{equation}
\label{eq3}
\ddot {e}_1 +\omega _{10}^2 e_1 +\delta _1 \dot {e}_1 =-\frac{p_{\rm{ex}} 
}{\rho _0 R_{10} }-\frac{R_{20}^2 }{R_{10} D}\ddot {e}_2 ,
\end{equation}
and exchanging 1 and 2 (or 10 and 20) in the subscripts in this equation 
yields that for bubble 2,
\begin{equation}
\label{eq4}
\ddot {e}_2 +\omega _{20}^2 e_2 +\delta _2 \dot {e}_2 =-\frac{p_{\rm{ex}} 
}{\rho _0 R_{20} }-\frac{R_{10}^2 }{R_{20} D}\ddot {e}_1 .
\end{equation}
This kind of system of differential equations is called a (linear) coupled 
oscillator model or a self-consistent model,\cite{ref28,ref29} and is known to 
be third-order accuracy with respect to $1/D$.\cite{ref30} The same (or 
essentially 
the same) formula has been employed in several studies considering acoustic 
properties of coupled bubbles.\cite{ref4,ref12,ref28,ref29,ref30,ref31,ref32}

Shima\cite{ref12} and Zabolotskaya,\cite{ref4} assuming $\delta _j \approx 0$ (for $j=1$ 
and 2), derived the theoretical formula for the natural frequencies, 
represented as
\begin{equation}
\label{eq5}
(\omega _{10}^2 -\omega ^2)(\omega _{20}^2 -\omega ^2)-\frac{R_{10} R_{20} 
}{D^2}\omega ^4=0.
\end{equation}
This equation predicts the existence of up to two natural frequencies (or 
eigenvalues of the system (\ref{eq3}) and (\ref{eq4})) per bubble. Exchanging 10 and 20 in 
the subscripts in this equation yields the same equation; namely, when 
$\delta _j \approx 0$, both the bubbles have the same natural frequencies. 
The higher and the lower natural frequencies reveal an upward and a downward 
shift, respectively, as $D$ decreases.\cite{ref12,ref13}

The theoretical formula for the transition frequencies, derived by 
Ida,\cite{ref13} can be obtained based on the harmonic steady-state solution of 
system (\ref{eq3}) and (\ref{eq4}). Assuming $p_{\rm{ex}} =-P_a \sin \omega t$, the solution is determined as follows:
\begin{equation}
\label{eq6}
e_1 =K_1 \sin (\omega t-\phi _1 ),
\end{equation}
where
\begin{equation}
\label{eq7}
K_1 =\frac{P_a }{R_{10} \rho _0 }\sqrt {A_1^2 +B_1^2 } ,
\end{equation}
\begin{equation}
\label{eq8}
\phi _1 =\tan ^{-1}\left( {\frac{B_1 }{A_1 }} \right)
\end{equation}
with
\[
A_1 =\frac{H_1 F+M_2 G}{F^2+G^2},
\quad
B_1 =\frac{H_1 G-M_2 F}{F^2+G^2},
\]
\[
F=L_1 L_2 -\frac{R_{10} R_{20} }{D^2}\omega ^4-M_1 M_2 ,
\]
\[
G=L_1 M_2 +L_2 M_1 ,
\quad
H_1 =L_2 +\frac{R_{20} }{D}\omega ^2,
\]
\[
L_1 = \omega _{10}^2 -\omega ^2,
\quad
L_2 = \omega _{20}^2 -\omega ^2,
\]
\[
M_1 =\delta _1 \omega ,
\quad
M_2 =\delta _2 \omega .
\]
Exchanging 1 and 2 (or 10 and 20) in the subscripts in these equations 
yields the formula for bubble 2. Based on the definition, the transition 
frequencies are given by\cite{ref13}
\begin{equation}
\label{eq9}
H_1 F+M_2 G=0.
\end{equation}
This equation predicts the existence of up to three transition frequencies 
per bubble\cite{ref13}; this number is greater than that of the natural frequencies 
given by eq.~(\ref{eq5}). This result means that in a multibubble case, the 
phase shift can take place not only around the bubbles' natural frequencies but 
also around some other frequencies. Moreover, it was shown in 
ref.~\citen{ref13} that 
the transition frequencies other than the natural frequencies do not cause 
the resonance response; namely, one of the three transition frequencies has 
different physical meanings from those of the natural frequency.

\begin{figure}
\begin{center}
\includegraphics[width=10cm]{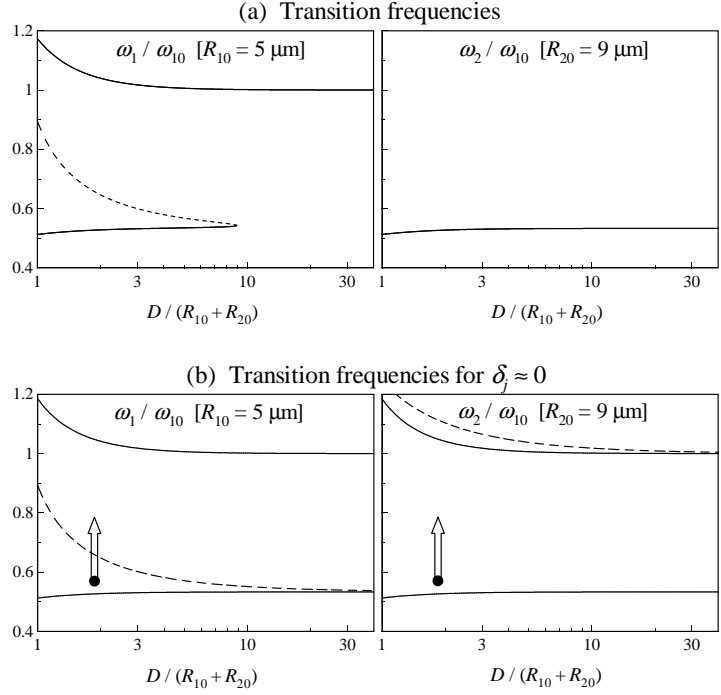}
\caption{
Transition frequencies as functions of the distance between the bubbles. The lower figure shows the transition frequencies in the case where the damping effect is neglected. The dashed lines denote the transition 
frequencies that do not cause resonance. The arrows are used for a 
discussion described in the text.}
\label{fig1}
\end{center}
\end{figure}

Figure \ref{fig1}(a) shows the transition frequencies of bubbles 1 and 2, $\omega _1 
$ and $\omega _2 $, for $R_{10} =5$ $\mu$m and $R_{20} =9$ $\mu$m, as 
functions of $D/(R_{10} +R_{20} )$. These bubble radii are chosen to be 
small enough so that the sphericity of the bubbles is maintained 
sufficiently. The parameters are set to $\rho _0 =1000$ kg/m$^{3}$, $\kappa 
=1.33$, $P_0 =1$ atm, and $\sigma =0.0728$ N/m. The damping factors are 
determined by the sum of the viscous and radiation ones as $\delta _j =(4 \mu
/\rho _0 R_{j0}^2 )+(\omega ^2R_{j0} /c)$ since the DNS technique does not 
consider the thermal conduction,\cite{ref18} where the viscosity of the liquid $\mu =1.137\times 10^{-3}$ kg/(m s) and its sound speed $c=1500$ m/s. As a 
reference, in Fig.~\ref{fig1}(b) we display the transition frequencies for $\delta _j 
\to 0$. As discussed previously,\cite{ref13} the highest and the second highest 
transition frequencies of the larger bubble tend to vanish when the damping 
effect is sufficiently strong; in the present case, they disappear 
completely. The second highest and the lowest transition frequencies of the 
smaller bubble cross and vanish at a certain distance, and only one 
transition frequency remains for sufficiently large $D$. For $D\to \infty $, 
the transition frequencies remaining converge to the partial natural 
frequencies of the corresponding bubble. The solid lines displayed in 
Fig.~\ref{fig1}(b) denote the transition frequencies that correspond to the natural 
frequencies given by eq.~(\ref{eq5}).

We clarify here physical meanings of the transition frequencies that do not 
accompany resonance, which meanings have not yet been described in the 
literature. Let us consider the bubbles under the condition indicated by the 
dots (the origin of the arrows) shown in Fig.~\ref{fig1}. Bubble 2 under this 
condition emits a strong sound (denoted below by $p_2 (D)$) whose oscillation 
phase is (almost) out-of-phase with the external sound $p_{{\rm ex}}$, because 
the driving frequency stays near and slightly above the natural frequency of 
this bubble. If $p_2 (D)$ is measured at a point sufficiently near bubble 2, 
its amplitude can be larger than that of $p_{{\rm ex}}$ and hence the phase of 
the total sound pressure, $p_{{\rm ex}}  + p_2 (D)$, may be almost the same as 
that of $p_2 (D)$, \textit{i.e.}, almost out-of-phase with $p_{{\rm ex}}$. As 
the driving frequency shifts toward a higher range along the arrows, the 
absolute value of $p_2 (D)$ decreases and, at a certain frequency, becomes 
lower than $\left| {p_{{\rm ex}} } \right|$; the phase of the total sound 
pressure finally becomes (almost) in-phase with $p_{{\rm ex}} $. This 
transition of the power balance of the two sounds results in the phase reversal 
of bubble 1 without accompanying the resonance response.

\subsection{The secondary Bjerknes force}
The secondary Bjerknes force acting between two pulsating bubbles is 
represented by\cite{ref1,ref2,ref3}
\begin{equation}
\label{eq10}
{\rm {\bf F}}\propto \langle {\dot {V}_1 \dot {V}_2 } \rangle 
\frac{{\rm {\bf r}}_2 -{\rm {\bf r}}_1 }{\left\| {{\rm {\bf r}}_2 -{\rm {\bf 
r}}_1 } \right\|^3},
\end{equation}
where $V_j $ and ${\rm {\bf r}}_j $ denote the volume and the position 
vector, respectively, of bubble $j$, and $\langle \cdots \rangle 
$ denotes the time average. Using eqs.~(\ref{eq6})--(\ref{eq8}), this equation can be rewritten as\cite{ref1}
\begin{equation}
\label{eq11}
{\rm {\bf F}}\propto K_1 K_2 \cos (\phi _1 -\phi _2 )\frac{{\rm {\bf r}}_2 
-{\rm {\bf r}}_1 }{\left\| {{\rm {\bf r}}_2 -{\rm {\bf r}}_1 } \right\|^3}.
\end{equation}
The sign reversal of this force occurs only when the sign of $\cos (\phi _1 
-\phi _2 )$ (or of $\langle {\dot {V}_1 \dot {V}_2 } \rangle )$ 
changes, because $K_1 >0$ and $K_2 >0$. Namely, the phase property of the 
bubbles plays an important role in the determination of the sign. Roughly 
speaking, the force is attractive when the bubbles pulsate in-phase with 
each other, while it is repulsive when they pulsate out-of-phase.

\begin{figure}
\begin{center}
\includegraphics[width=7.5cm]{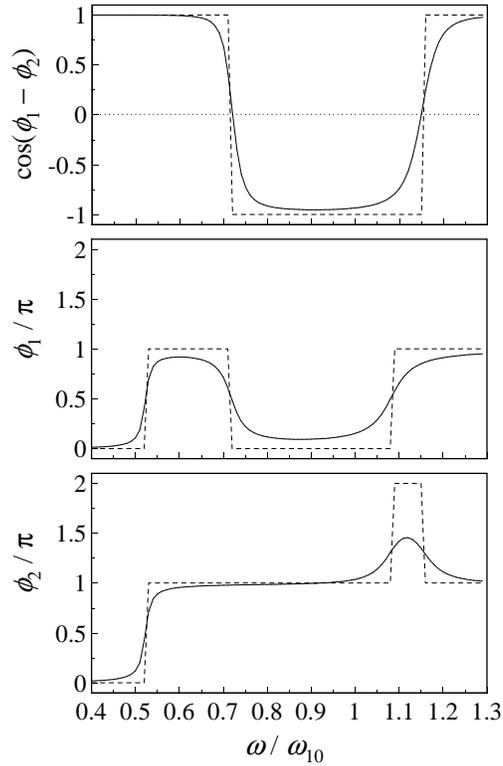}
\caption{Sign of the secondary Bjerknes force [$\cos (\phi _1 - \phi _2)$] and phase differences [$\phi _1$ and $\phi _2$], determined theoretically, as functions of the driving angular frequency. The positive value of $\cos (\phi _1 - \phi _2)$ indicates the attraction, while the negative one indicates the repulsion. The dashed lines denote the results 
for $\delta _j \approx 0$.}
\label{fig2}
\end{center}
\end{figure}

The solid lines displayed in Fig.~\ref{fig2} denote 
$\cos (\phi _1 -\phi _2 )$, $\phi 
_1 $, and $\phi _2 $ as functions of $\omega /\omega _{10} $, and the dashed 
lines denote those for $\delta _j \approx 0$. The physical parameters are 
the same as used for Fig.~\ref{fig1}, except for the separation distance fixed 
to $D=20$ $\mu$m [$D/(R_{10} +R_{20} )\approx 1.43]$. In this figure, we can 
observe sign reversals of $\cos (\phi _1 -\phi _2 )$ at $\omega /\omega 
_{10} \approx 0.72$ and $\omega /\omega _{10} \approx 1.15$. In 
ref.~\citen{ref14}, 
it was shown theoretically that both the sign reversals are due to the 
transition frequencies that do \textit{not} correspond to the natural frequencies; that 
is, the sign reversals take place near (or, when $\delta _j \to 0$, at) 
those frequencies. The present result follows that theoretical prediction. 
The respective reversals are observed near the second highest transition 
frequency of bubble 1 and near the highest of bubble 2 for $\delta _j \to 
0$. Meanwhile, we can observe $\phi _2 >\pi $, which should not be observed 
in a single-bubble case. Ida\cite{ref14} explained that such a large phase delay is 
realizable in a multibubble case, by the radiative interaction.

We make here a remark regarding the estimation of the points at which the 
sign reversal of the force takes place. It was shown previously\cite{ref14} 
that the 
transition points of the force are hardly changed by the damping effects, 
even when bubbles are small ($R_{j0} \sim 1$ $\mu$m) or relatively large 
($R_{j0} \sim 1$ mm). The present result displayed in Fig.~\ref{fig2} follows 
that 
finding. These results may allow us to consider that the simple theoretical 
formulas for the transition frequencies not causing resonance,
\[
\omega _1^2 =\frac{\omega _{20}^2 }{1-R_{20} /D}\quad \mbox{and}\quad \omega 
_2^2 =\frac{\omega _{10}^2 }{1-R_{10} /D}
\]
derived for $\delta _1 \to 0$ and $\delta _2 \to 0$,\cite{ref13} can be a good 
approximation of the transition points, even in the cases of $\delta _j \ne 
0$. These formulas yield $(\omega _1 /\omega _{10} )=0.719$ and $(\omega _2 
/\omega _{10} )=1.155$, which are consistent with the result shown in 
Fig.~\ref{fig2}.

\section{Numerical results and discussion}
\label{sec3}
In this section, the DNS technique\cite{ref18,ref19} is employed to verify the 
theoretical results for the transition frequencies. The governing equations 
are the compressible Navier--Stokes equations
\begin{eqnarray*}
&& \frac{\partial \rho }{\partial \,t}+{\rm {\bf u}}\cdot \nabla \rho =-\rho 
\nabla \cdot {\rm {\bf u}}, \\
&& \frac{\partial {\rm {\bf u}}}{\partial t}+{\rm {\bf u}}\cdot \nabla {\rm 
{\bf u}}=-\frac{\nabla p}{\rho }+\frac{1}{\rho }\left( {2\nabla \cdot (\mu 
{\rm {\bf T}})-\frac{2}{3}\nabla (\mu \nabla \cdot {\rm {\bf u}})} 
\right)+\frac{{\rm {\bf F}}_{st} }{\rho }, \\
&& \frac{\partial p}{\partial \,t}+{\rm {\bf u}}\cdot \nabla p=-\rho C_S^2 
\nabla \cdot {\rm {\bf u}},
\end{eqnarray*}
where $\rho $, ${\rm {\bf u}}$, $p$, ${\rm {\bf T}}$, $\mu $, ${\rm {\bf 
F}}_{st} $, and $C_S $ denote the density, the velocity vector, the 
pressure, the deformation tensor, the viscosity coefficient, the surface 
tension as a volume force, and the local sound speed, respectively. This 
system of equations is solved over the whole computational domain divided by 
grids. The materials (the bubbles and a liquid surrounding them) moving on 
the computational grids are identified by a scalar function\cite{ref19} 
depending 
on a pure advection equation whose characteristic velocity is ${\rm {\bf 
u}}$. The bubbles' radii and the initial center-to-center distance between 
them are set by using the same values as used for Fig.~\ref{fig2}, that is, 
$R_{10} 
=5$ $\mu$m, $R_{20} =9$ $\mu$m, and $D(t=0)=20$ $\mu$m. The content inside 
the bubbles is assumed to be an ideal gas with a specific heat ratio of 
1.33, an equilibrium density of 1.23 kg/m$^{3}$, and a viscosity of 
$1.78\times 10^{-5}$ kg/(m s). The surrounding liquid is water, whose sound 
speed is determined by $C_S =\sqrt {7(p+3172.04P_0 )/\rho } $ with $p$ and 
$\rho $ being the local pressure and density, respectively. The other 
parameters are set to the same values as those used previously. The 
axisymmetric coordinate ($r,z)$ is selected for the computational domain, 
and the mass centers of the bubbles are located on the central axis of the 
coordinate. The grid widths are set to be constants as $\Delta r=\Delta 
z=0.25$ $\mu$m, and the numbers of them in the $r$ and the $z$ coordinates 
are 100 and 320, respectively. The sound pressure, applied as the boundary 
condition to the pressure, is assumed to be in the form of $p_{\rm{ex}} 
=P_a \sin \omega t$, where the amplitude $P_a $ is fixed to $0.3P_0 $ and 
the driving frequency is selected from the frequency range around the 
bubbles' partial natural frequencies. (For the sound amplitude assumed, 
nonlinear effects may not completely be neglected especially near the 
natural frequencies. However, we can, unfortunately, not use a lower sound 
pressure because of a numerical problem called 
``parasitic currents,''\cite{refa1,refa2} which rise to the surface when a 
small bubble or drop is in (nearly) 
steady state where the amplitudes of volume and shape oscillations are very 
low. Effects of nonlinearity are briefly discussed later.) The boundary 
condition for the velocity is free.

\begin{figure*}
\begin{center}
\includegraphics[width=7.5cm]{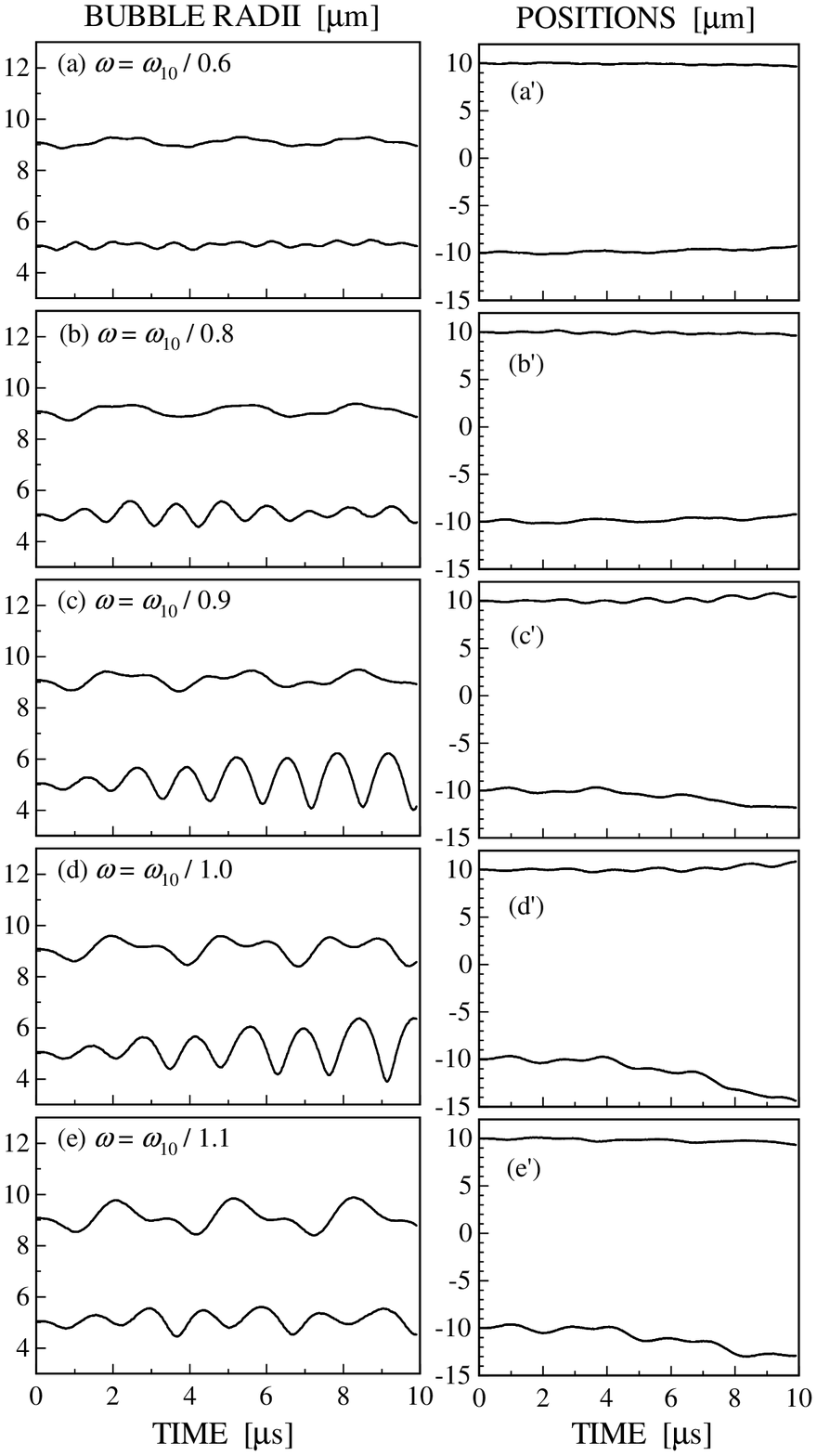}\quad
\includegraphics[width=7.5cm]{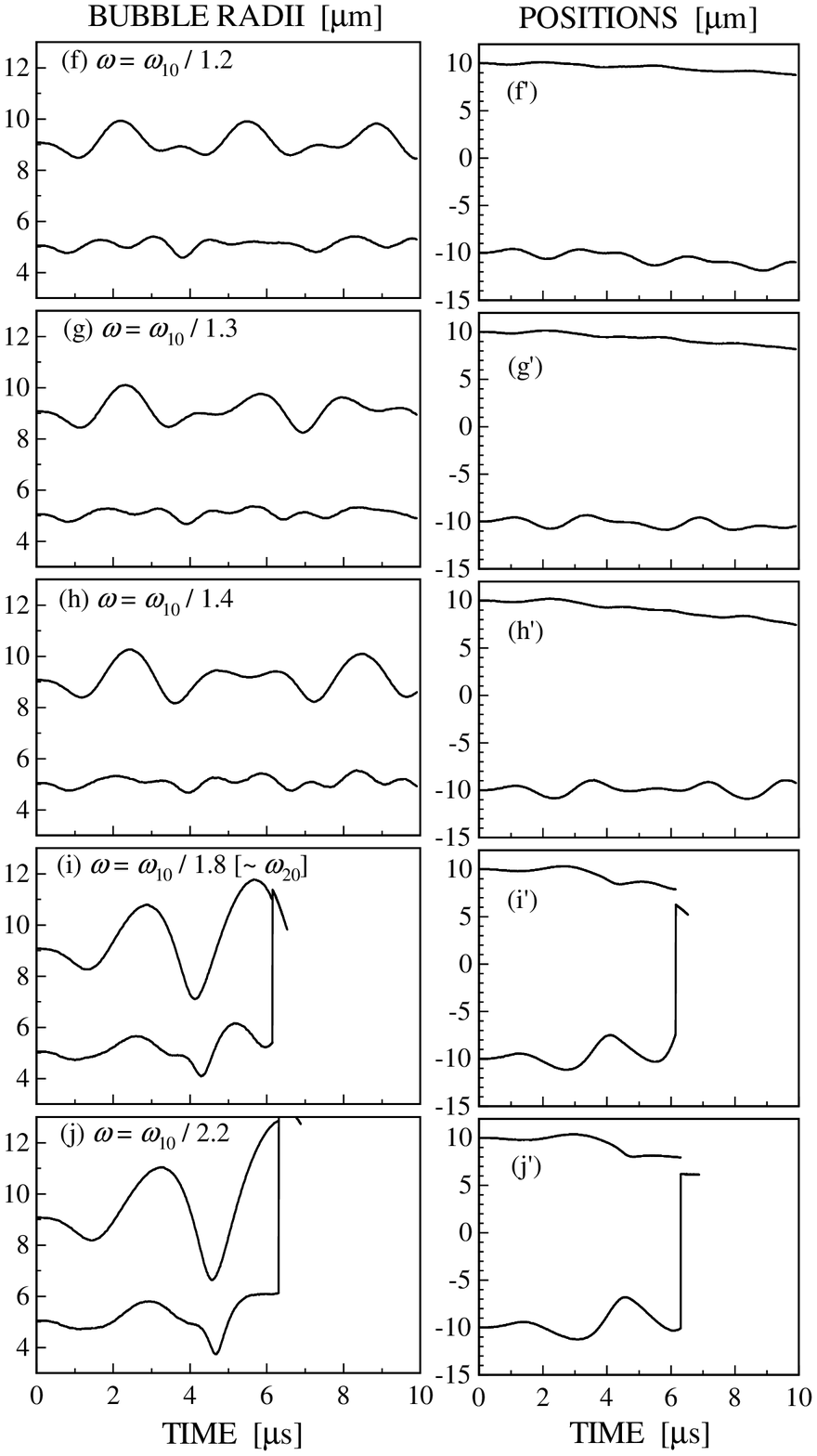}
\caption{DNS results: Bubble radii [(a) $\sim $ (j)] and corresponding 
positions [(a') $\sim$ (j')] as functions of time for different driving 
frequencies. The lower lines in (a')--(j') denote the position of the 
smaller bubble. The bubbles coalesce at the time where the number of lines 
becomes one, see panels (i), (i'), (j), and (j').}
\label{fig3}
\end{center}
\end{figure*}

\begin{figure}
\begin{center}
\includegraphics[width=7.5cm]{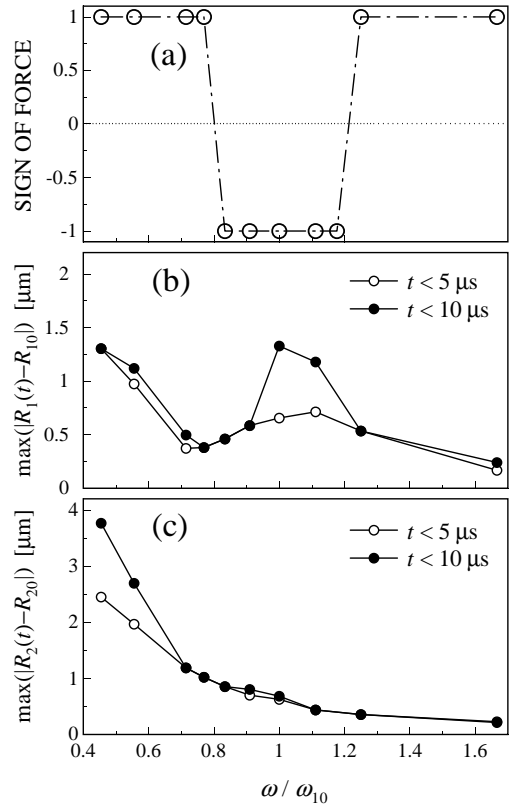}
\caption{DNS results: (a) Sign of the force, (b) pulsation amplitude of bubble 
1, and (c) that of bubble 2, as functions of the driving frequency. The 
amplitudes were measured for $t < 5$ $\mu$s ($\circ$), and for $t < 10$ $\mu$s but until the coalescence has been observed ($\bullet$). 
The result for $\omega = \omega _{10} /0.85$, not shown in Fig.~\ref{fig3}, is presented in (a); see also Fig.~\ref{fig5}.}
\label{fig4}
\end{center}
\end{figure}

\begin{figure}
\begin{center}
\includegraphics[width=6.8cm]{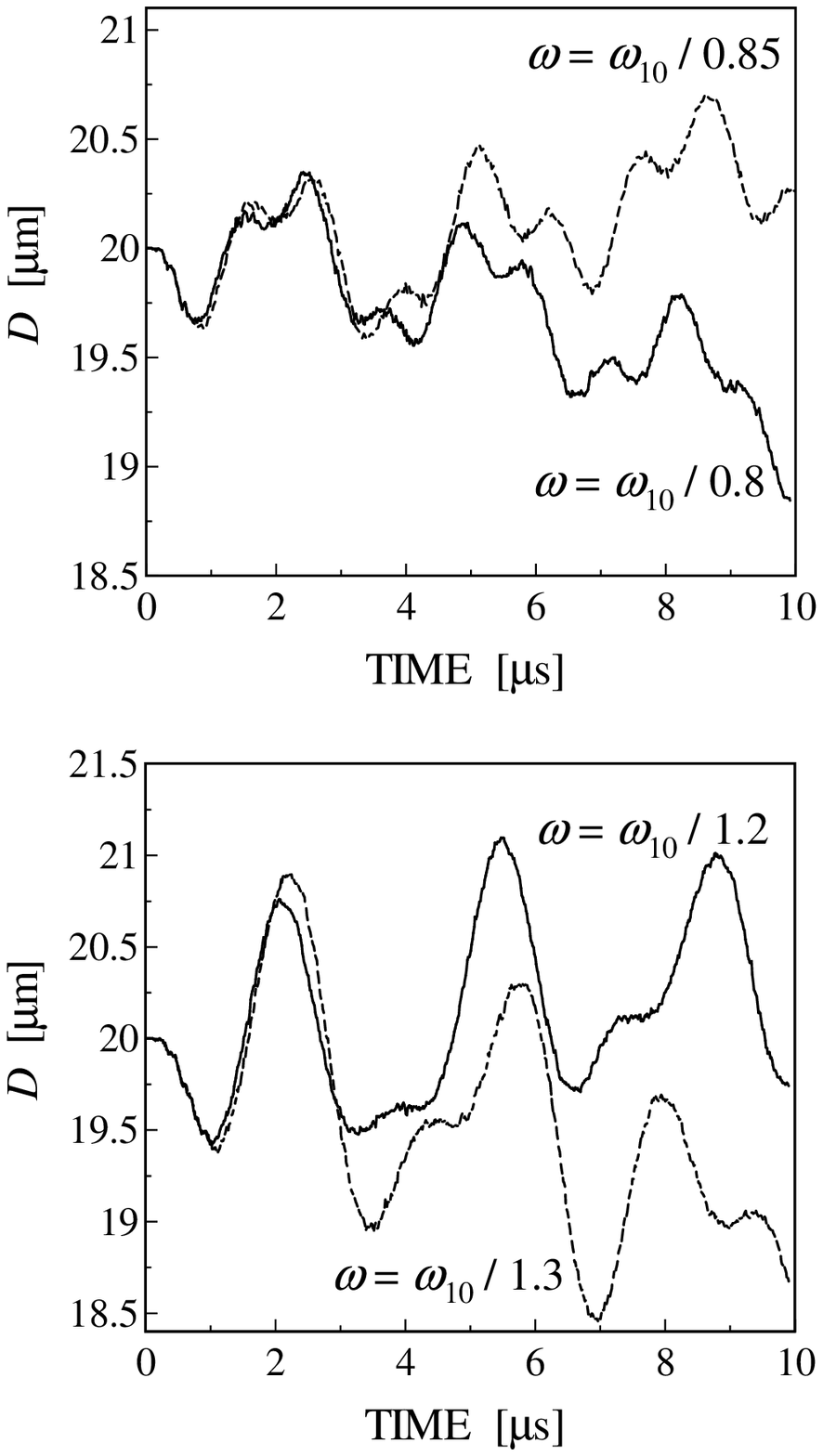}
\caption{Time-dependent distances between the mass centers of the bubbles in 
the cases where the deviation in the distance is small (for 
$\omega  = \omega _{10} /0.8$, $\omega _{10} /0.85$, $\omega _{10} /1.2$, and $\omega _{10} /1.3$). Note that the result for $\omega  = \omega _{10} /0.85$, not shown in Fig.~\ref{fig3}, is added. [The low-amplitude, 
high-frequency oscillations observed in these curves may be due to a 
numerical error originating in the calculation of such a small deviation 
(comparable to the grid width) on a discrete computational domain.]}
\label{fig5}
\end{center}
\end{figure}

Figure \ref{fig3} displays the bubbles' (mean) radii and mass centers as 
functions of time for different $\omega $, and figure \ref{fig4} displays the 
sign of the force 
(a), determined by observing the direction of the bubbles' translation, and 
the bubbles' pulsation amplitudes (b and c). From these figures, we know 
that the smaller bubble has two resonance frequencies, one at $\omega 
\approx \omega _{10} /0.9$ ($\approx 1.1\omega _{10} )$ and the other at 
$\omega \approx \omega _{10} /2.2$ ($\approx 0.45\omega _{10} )$, though the 
former seems to decrease with time because of the repulsion of the bubbles 
(See Fig.~\ref{fig1}, which reveals that the highest transition frequency of 
bubble 
1, causing resonance, decreases as $D$ increases). The former resonance 
frequency is higher than $\omega _{10} $, and the latter is lower than 
$\omega _{20} $ ($\approx 0.53\omega _{10} )$. These respective resonances 
are obviously due to the highest and the lowest transition frequencies of 
bubble 1, both of which correspond to the natural frequencies. The same 
figure shows that the larger bubble may have a resonance frequency, at 
$\omega \approx \omega _{10} /2.2$.

The sign of the interaction force changed twice in the frequency region 
considered. In the region between $\omega =\omega _{10} /0.8$ and $\omega 
_{10} /0.9$, being near to but above the higher resonance frequency of the 
smaller bubble discussed above, the attractive force turns into repulsion as 
$\omega $ decreases, and, at $\omega \approx \omega _{10} /1.2$ ($\approx 
0.83\omega _{10} )$ the repulsive force turns back into attraction. (See 
also Fig.~\ref{fig5}, which shows $D(t)$ in the cases where the deviation in it is 
small.) It may be difficult to say, using only these numerical results, that 
the former reversal is not due to the higher natural frequency of the 
smaller bubble, because the reversal took place near it and the highest 
transition frequency of the larger bubble is close to it. Therefore, in the 
following we will focus our attention on the latter sign reversal, which 
occurred at $\omega $ between $\omega _{10} $ and $\omega _{20} $.

\begin{figure}
\begin{center}
\includegraphics[width=7.8cm]{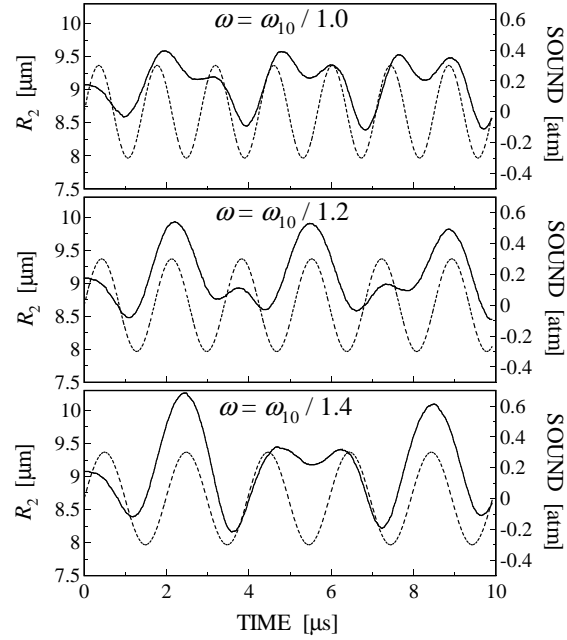}
\caption{Radius of the larger bubble (solid lines) and sound pressure as functions of time, for around $\omega  = \omega _{10} /1.2$.}
\label{fig6}
\end{center}
\end{figure}

The latter reversal indicates that a kind of characteristic frequency should 
exist in the frequency region between the partial natural frequencies of the 
bubbles. It is evident that this characteristic frequency is not the 
resonance frequency of the larger bubble, which is, as already discussed, 
much lower. This result is in opposition to the assumption described by 
Doinikov and Zavtrak.\cite{ref5,ref6} Also, the theory for the natural 
frequencies (eq.~(\ref{eq5})) cannot explain this reversal, because it predicts 
no natural 
frequency in the frequency region between $\omega _{10} $ and $\omega _{20} 
$. This characteristic frequency is, arguably, the second highest transition 
frequency of the smaller bubble, as was predicted by Ida.\cite{ref14} As was 
proved theoretically in ref.~\citen{ref13}, resonance response is not observed 
around this 
characteristic frequency. In order to confirm that this characteristic 
frequency is not of the larger bubble, we display in Fig.~\ref{fig6} the 
$R_2$--time and $p_{\rm{ex}}$--time curves for the area around $\omega 
=\omega _{10} /1.2$. This figure shows clearly that the pulsation phase of 
the larger bubble does not reverse in this frequency region; the bubble 
maintains its out-of-phase pulsation with the external sound (\textit{i.e.}, the bubble's 
radius is large when the sound amplitude is positive), although other modes, 
which may result from the transient, appear.

\begin{figure}
\begin{center}
\includegraphics[width=7cm]{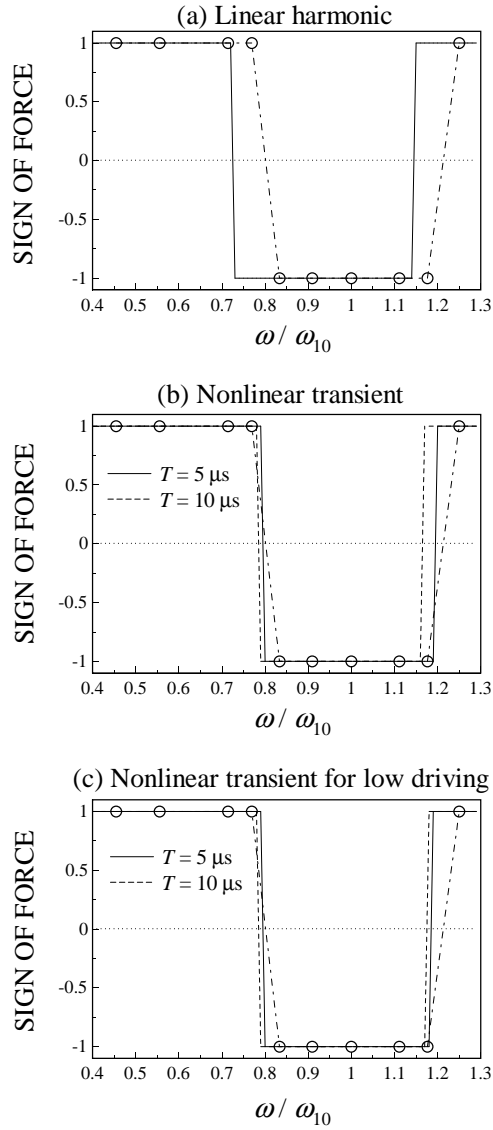}
\caption{Comparison of the DNS results with (a) the linear harmonic solution, (b) the nonlinear numerical results for $P_a  = 0.3 P_0$, and (c) those for $P_a  = 0.001 P_0$. The circles denote the DNS results, where the 
positive value denotes the attraction while the negative value denotes the 
repulsion.}
\label{fig7}
\end{center}
\end{figure}

\begin{figure}
\begin{center}
\includegraphics[width=7.5cm]{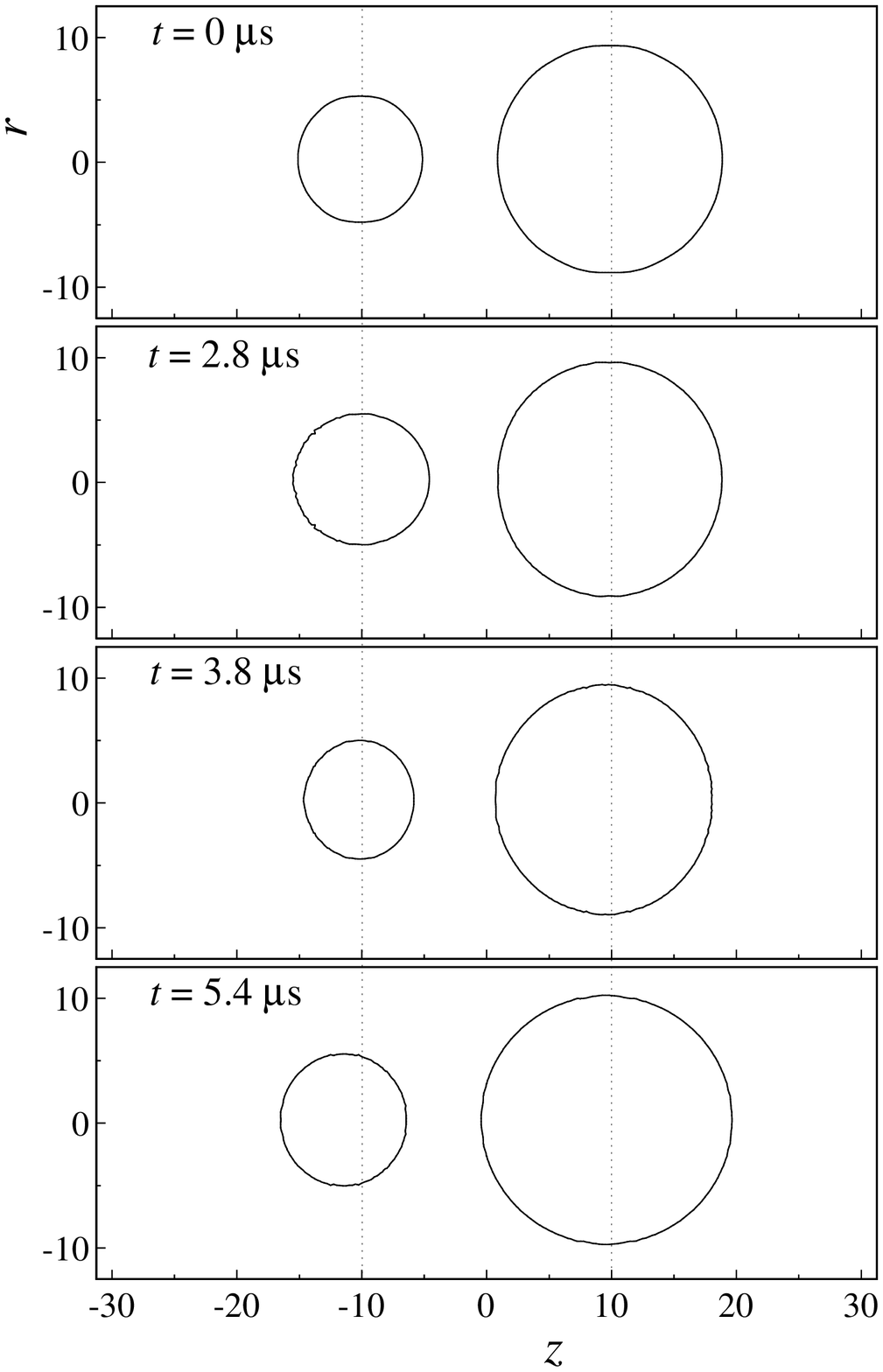}
\caption{Bubble surfaces for $\omega  = \omega _{10} /1.2$ at selected times, given by the DNS technique. Similar figures for $\omega  = \omega _{10}$ and $\omega  = \omega _{10} /1.8$ can be found in our previous paper \cite{ref18}.}
\label{fig8}
\end{center}
\end{figure}

Here, we discuss how this transient and a nonlinear effect act on the 
quantitative nature of the sign distribution. In Fig.~\ref{fig7}(a), we show 
the sign of the force as functions of $\omega $, determined by the linear 
theory (the solid line) and by the DNS (the circles). The positive value 
denotes the attraction, while the negative value the repulsion. A noticeable 
quantitative discrepancy can be observed between these results, whereas they 
are in agreement in a qualitative sense. We attempt here to identify what 
caused this discrepancy, using a nonlinear model. Mettin \textit{et 
al}.~\cite{ref33} proposed a nonlinear coupled oscillator model for a 
double-bubble system,
\begin{equation}
\label{eq12}
\left( {1-\frac{\dot {R}_1 }{c}} \right)R_1 \ddot {R}_1 +\left( 
{\frac{3}{2}-\frac{\dot {R}_1 }{2c}} \right)\dot {R}_1^2 =\frac{1}{\rho_0 
}\left( {1+\frac{\dot {R}_1 }{c}} \right)p_{\rm{k}\,1} +\frac{R_1 }{\rho_0 
c}\frac{dp_{\rm{k}\;1} }{dt}-\frac{1}{D}\frac{d}{dt}(\dot {R}_2 R_2^2 ),
\end{equation}
\begin{equation}
\label{eq13}
\left( {1-\frac{\dot {R}_2 }{c}} \right)R_2 \ddot {R}_2 +\left( 
{\frac{3}{2}-\frac{\dot {R}_2 }{2c}} \right)\dot {R}_2^2 =\frac{1}{\rho_0 
}\left( {1+\frac{\dot {R}_2 }{c}} \right)p_{\rm{k}\,2} +\frac{R_2 }{\rho_0 
c}\frac{dp_{\rm{k}\,2} }{dt}-\frac{1}{D}\frac{d}{dt}(\dot {R}_1 R_1^2 ),
\end{equation}
with
\[
p_{\rm{k}\,j} =\left( {P_0 +\frac{2\sigma }{R_{j0} }} \right)\left( 
{\frac{R_{j0} }{R_j }} \right)^{3\kappa }-\frac{2\sigma }{R_j }-\frac{4\mu 
\dot {R}_j }{R_j }-P_0 -p_{\rm{ex}} \quad \mbox{for}\;j=1\mbox{ or }2,
\]
based on the Keller--Miksis model\cite{ref34} taking into account the 
viscosity and compressibility of the surrounding liquid with first-order 
accuracy. In this system of equations, the last terms of eqs.~(\ref{eq12}) and (\ref{eq13}) 
represent the radiative interaction between the bubbles. Using this model, 
$R_j $ and $\dot {R}_j $ are calculated by the fourth-order Runge--Kutta method, and subsequently the time average in eq.~(\ref{eq10}) is performed 
to determine the sign of the force. Although the quantitative accuracy of 
this model may not be guaranteed for a small $D$,\cite{ref33} a rough 
estimation of 
the influences of the transient and nonlinearity might be achieved. The 
physical parameters are the same as used for Fig.~\ref{fig1}, and $D$ is fixed 
to 
$20$ $\mu$m, that is, the translational motion is neglected. The solid and the 
dashed lines displayed in Fig.~\ref{fig7}(b) show
\[
G(T)\equiv {\rm sgn}\left[ {{\int_0^T {\dot {V}_1 \dot {V}_2 dt} } \mathord{\left/ 
{\vphantom {{\int_0^T {\dot {V}_1 \dot {V}_2 dt} } T}} \right. 
\kern-\nulldelimiterspace} T} \right]
\]
for $T=5$ $\mu$s and 10 $\mu$s, respectively, given using the nonlinear 
model, where ${\rm sgn}[f]=1$ for $f>0$ and ${\rm sgn}[f]=-1$ otherwise. These numerical 
results are in good agreement with the DNS results (the circles). Figure 
7(c) shows the numerical results given using the nonlinear model with a 
very low driving pressure ($P_a =0.001P_0 )$. These results are also in good 
agreement with the DNS results, proving that, in the present case, the 
nonlinearity in the pulsations is not dominant for the sign reversals. 
Furthermore, it can be proved that the bubbles' shape oscillation is also not 
dominant for the sign reversal. Figure \ref{fig8} shows the bubble surfaces for 
$\omega =\omega _{10} /1.2$ at selected times. Only a small deformation of 
the bubbles can be found in this figure. (The numerical results provided 
previously in ref.~\citen{ref18}, given using the same DNS technique, reveal 
that the bubbles' sphericity is well maintained even for 
$\omega =\omega _{10} $, whereas a noticeable deformation is observed for 
$\omega \approx \omega _{20} $.) These results allow us to consider that the 
transient is one of the dominant origins of the noticeable quantitative 
discrepancy found in Fig.~\ref{fig7}(a).

There are some other physical factors that may also be able to cause the 
quantitative discrepancy but are not taken into account in both the linear and 
nonlinear coupled 
oscillator models. It is known, for example, that the translational motion of 
bubbles can alter the bubbles' pulsation. As has been proved by several 
researchers,\cite{ref35,ref36,ref30} if the translational motion is taken into 
consideration in deriving a theoretical model, high-order terms appear in the 
equations of radial motion, in which not only the bubbles' radii but also their 
translational velocities are involved. Also, in certain cases the viscosity of 
the surrounding liquid can alter the magnitude and sign of the interaction 
force, because acoustic streaming is induced around the bubbles.\cite{refaa1} 
Investigating the influences of those factors on the quantitative discrepancy 
using a higher-order model would be an interesting and important subject.

\section{Concluding remarks}
\label{sec4}
In summary, we have verified the recent theoretical results regarding the 
transition frequencies of two acoustically interacting bubbles\cite{ref13} and 
the sign reversal of the secondary Bjerknes force,\cite{ref14} using a DNS 
technique.\cite{ref18,ref19} The present 
numerical results, given by the DNS technique, support those theoretical 
results at least in a qualitative sense. The most important point validated 
by DNS is that at least the sign reversal occurring when the driving 
frequency stays between the bubbles' partial natural frequencies is 
obviously not due to the natural (or the resonance) frequencies of the 
double-bubble system. This conclusion is in opposition to the previous 
explanation described by Doinikov and Zavtrak,\cite{ref5,ref6} but is 
consistent with the most recent interpretation by Ida\cite{ref14} described 
based on analyses of the transition frequencies,\cite{ref13,ref15} thus 
validating the assertion that the transition frequencies not corresponding to 
the natural frequencies exist and that the notion ``transition frequency'' is 
useful for understanding the sign reversal of the force.

\section*{Acknowledgment}
This work was supported by the Ministry of Education, Culture, Sports, 
Science, and Technology of Japan (Monbu-Kagaku-Sho) under an IT research 
program ``Frontier Simulation Software for Industrial Science.''

\end{document}